\documentclass[twocolumn,showpacs,preprintnumbers,amsmath,amssymb,aps]{revtex4} 

\usepackage{epsfig,amsopn}
\usepackage{graphicx}
\usepackage{amsmath,amssymb}
\usepackage{amsthm}
\usepackage{enumerate}
\usepackage{ctable}
\usepackage{hyperref}

\newcommand\bea{\begin{eqnarray}}
\newcommand\eea{\end{eqnarray}}
\newcommand\beq{\begin{equation}}
\newcommand\eeq{\end{equation}}

\def\qp{\Psi^{qp}}
\def\ef{\Psi^e}

\definecolor{dgreen}{rgb}{0.0,0.36,0.0}
\definecolor{dred}{rgb}{0.7,0.1,0.1}
\definecolor{dblue}{rgb}{0.1,0.1,0.7}
 
\begin{document}

\title{Thermoelectric probe for neutral edge modes in the fractional quantum Hall regime}

\author{Giovanni Viola$^{1}$, Sourin Das$^{2}$, Eytan Grosfeld$^{3}$ and Ady Stern$^{1}$ }
\affiliation{$^{1}$ Department of Condensed Matter Physics, Weizmann Institute of Science, Rehovot 76100, Israel \\
$^{2}$ Department of Physics and Astrophysics, University of Delhi, Delhi 110007, India \\
$^{3}$ Department of Physics, Ben-Gurion University of the Negev, Beer Sheva 84105, Israel}

\date{August 26, 2012}

\begin{abstract}
The  {$\nu=5/2$} anti-Pfaffian state and the {$\nu=2/3$} state are believed to have an edge composed of counter-propagating charge and neutral modes. This situation allows the generation of a pure thermal bias between two composite edge states across a quantum point contact (QPC) as was experimentally established in Nature {\bf 466}, 585 (2010). We show that replacing the QPC by a quantum dot provides a natural way for detecting the neutral modes via the DC current generated by the thermoelectric response of the dot. We also show that the degeneracies of the dot spectrum, dictated by the conformal field theories (CFTs) describing these states, induce asymmetries in the thermoelectric current peaks. This in turn provides a direct fingerprint of the corresponding CFT.
\end{abstract}
\maketitle

{\it{Introduction}:---}It has been thirty years since the discovery of the fractional quantum Hall effect (FQHE)~\cite{FQHE}, yet even now this remarkable quantum regime continues to give rise to some of the most fascinating collective behaviors of electrons. Of particular interest are the non-Abelian states~\cite{TQC}, whose underlying quasiparticles are predicted to obey non-Abelian braiding statistics~\cite{MooreRead,ReadGreen,Ivanov}. A prime candidate is the plateau at filling factor {$\nu=5/2$}, which may be described by the Pfaffian state~\cite{MooreRead}, or by its particle-hole conjugate, the anti-Pfaffian state (APF)~\cite{LevinHalperinRosenow,LeeRyuNayakFisher}. Neutral edge modes are present in both of these non-Abelian states, for the former flowing ``downstream'' along the direction of the chiral charge-carrying mode, and for the latter ``upstream'' in the opposite direction. The conformal field theories (CFTs) describing these edge states reflect the properties of the corresponding bulks, and contain signatures of their associated non-Abelian quasiparticles~\cite{TQC,MooreRead}.

Neutral edge modes are predicted to exist both in
Abelian and non-Abelian quantum Hall states. An exciting
development is the recent experimental detection of neutral
edge modes using noise measurements~\cite{BidOfek}. The ideas leading
to this breakthrough were theoretically established in two papers~\cite{FeldmanLi,GrosfeldDas}. Grosfeld and Das~\cite{GrosfeldDas} predicted that neutral edge modes can be thermally biased by electrical means
and then detected via noise measurements, while Feldman and Li~\cite{FeldmanLi}
considered a non-equilibrium coherent neutral beam, also leading
to an increase in noise. Due to the particular details of this measurement procedure, it is mostly sensitive to the presence of an {\em upstream} neutral mode. 
Hence, its detection for filling factor $\nu=2/3$ confirmed a long-standing prediction~\cite{KaneFisherPolchinski,KaneFisher}; while its detection at $\nu=5/2$ singles out the APF~\cite{LevinHalperinRosenow,LeeRyuNayakFisher} over the Pfaffian~\cite{MooreRead} and the 331 state~\cite{Halperin} as the more promising description of the state. However, the measurement extracts no details about the particular field theory associated with the neutral edge state.  

\begin{figure}[htp]
\centering
\includegraphics[scale=0.4]{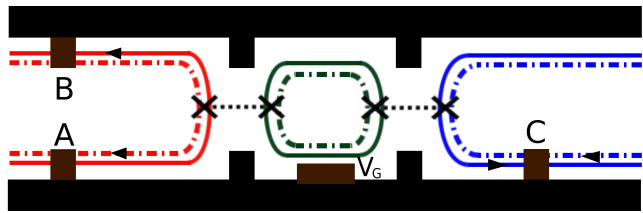}
\caption{A quantum dot defined within two QPCs in a Hall bar geometry.
The neutral (charge) modes are depicted by a dashed (solid) line. Charge
flows anticlockwise. A voltage bias is imposed on B and a net current is collected at C.}
\label{fig:setup}
\end{figure}

In this Letter we propose a scheme (see Fig.~\ref{fig:setup}) which can probe the presence and properties of an upstream flowing neutral edge using a thermoelectric measurement. A quantum Hall droplet is pinched using two quantum point contacts (QPCs) to form a quantum dot (QD). Two quantum Hall edge states, weakly coupled via electron tunneling to the left and right of the dot, are held at temperatures $T_L$ and $T_R$. This temperature difference is generated by pure electrical means ($V_B\neq 0$) and depends on the ability to pump energy upstream via the neutral mode~\cite{BidOfek,TakeiRosenow,DolevGross,GrossDolevMahalu,Yacoby,Pierre}, \emph{i.e.} $T_L\neq T_R$ (at $V_A=V_C=0$) signals the presence of the neutral mode.
We claim that this is a natural setup for a measurement of a thermoelectric effect. When an energy level of the dot gets slightly detuned from the chemical potential of the two reservoirs, electric current will flow through the dot since particle-hole (p-h) symmetry is broken. This results in a particle current emanating from the dot to one of the outgoing edge states and a hole current to the other (Eq.~(\ref{current2})). 

\begin{figure}[htp]
\centering
\includegraphics[scale=.58]{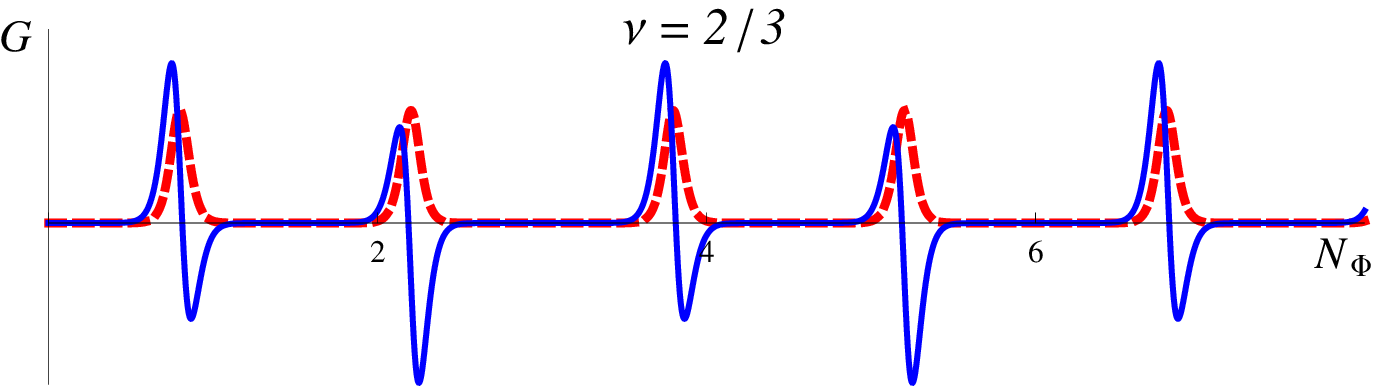}
\includegraphics[scale=.58]{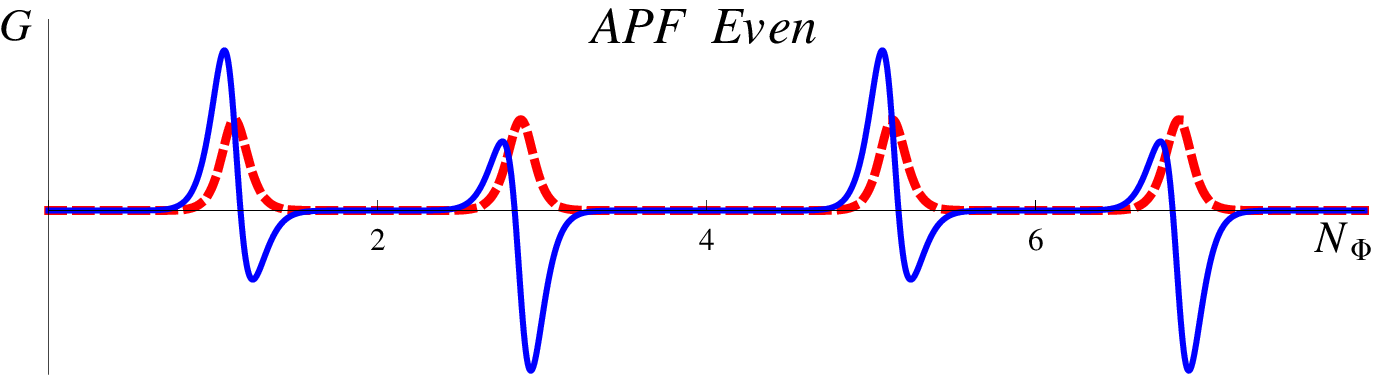}
\includegraphics[scale=.58]{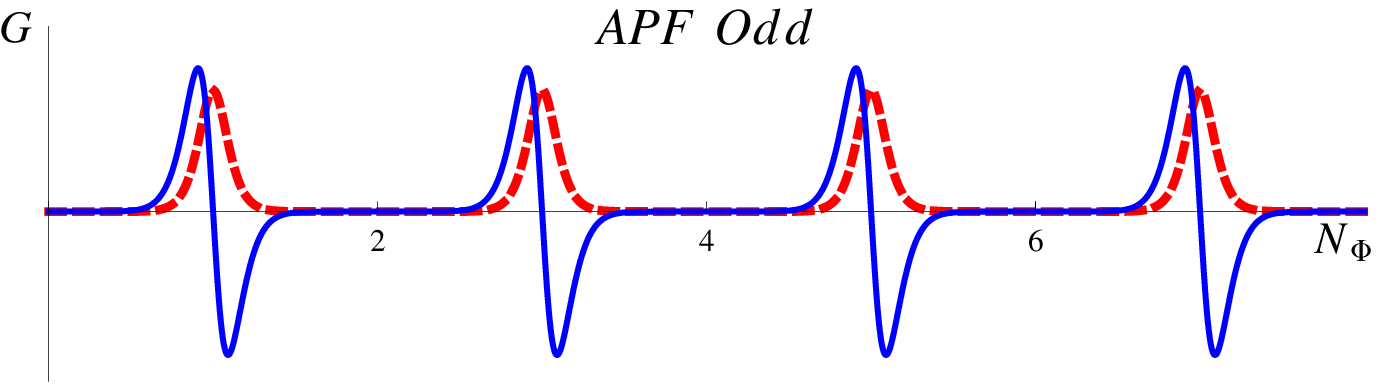}
\caption{The plots of the electric ($G$) and thermoelectric ($G_T$) coefficients in red-dashed and blue-solid lines as function of 
$N_{\Phi}=\frac{B(S-S_0)}{\phi_0}$, for  $\nu=2/3$ and the APF for both even and odd number of quasiparticles in the bulk. The temperature is set to be $T\simeq \frac{0.02 v_c}{R}$ and the neutral 
velocity is $v_n\simeq \frac{1}{6} v_c$. [$G$ and $G_T$ are plotted in identical arbitrary units]}
\label{fig:patterns}
\end{figure}

Furthermore, we show that the line shapes of such thermoelectric coulomb blockade peaks carry information about the spectrum of the edge state. The degeneracies of the resonant dot levels $\mathcal D_a$ (here $a=1,2,3,\ldots$ is a numbering of the peaks), which are a direct reflection of the CFT describing the edge states~\cite{CappelliZemba,Georgiev,CappelliViola}, induce asymmetries in the thermoelectric current (Fig.~\ref{fig:patterns}).
The degree of asymmetry increases with the degeneracy of the levels in the dot. The patterns of degeneracies are given by Eq.~(\ref{eq:deg23}) for $\nu=2/3$ and Eqs.~(\ref{eq:degAPFeven},\ref{eq:degAPFodd}) for $\nu=5/2$. Hence a measurement of the degeneracies via the asymmetries has the potential to sharply characterize these edge states.

{\it{Overview of the measurement scheme}:---}The main elements in the setup depicted in Fig.~\ref{fig:setup}  are: (i) the presence of a temperature gradient between left and right (L-R);  
and (ii) the presence of a QD with a given edge spectrum. 
Experimentally a thermal gradient that is not accompanied by a gradient in the electro-chemical potential can be created provided the neutral and the charge 
modes are counter-propagating~\cite{BidOfek,TakeiRosenow,DolevGross,GrossDolevMahalu,Yacoby}. The upstream neutral mode is heated up when (anti-clockwise) current is injected through a ohmic contact (B) and then fully absorbed by a nearby 
grounded terminal. The flow of energy in the opposite direction is translated to an increased temperature of the neutral and charge modes after full 
equilibration. 
The emergent temperature bias is therefore a direct fingerprint of the presence of an upstream mode in the external edge states. 
The role of the dot is to break the p-h symmetry of the system through the quantization of its energy levels. 
By controlling the position of the energy levels of the dot via a side gate ($V_G$), one can break p-h symmetry in a controlled manner hence 
generating a measurable current which switches sign as we sweep the level across the Fermi energy of the edge states.

When both L-R (via temperature bias) and p-h symmetries are broken thermoelectric response will ensue~\cite{Bergman}. The calculation of the thermoelectric effect proceeds in three stages:
First, we describe the theory of the edge state for $\nu=2/3$ and the Anti-Pfaffian $\nu=5/2$. Then we calculate the energy levels and degeneracies. 
Finally, we focus on a single (possibly degenerate) level and calculate the thermoelectric response through that level.

{\it{Theory of the edge state}:---} Here we review the essential features of the edge theories for the states we consider (a more detailed exposition is presented in the supplementary).
The $\nu=2/3$ state and the $\nu=5/2$ APF state are naively expected to present edge theories with counter-propagating charge modes~\cite{WenBook,KaneFisherPolchinski,Frohlich,LevinHalperinRosenow,LeeRyuNayakFisher,BisharaFieteNayak}. 
However, random impurities along the boundaries of the sample introduce RG-relevant inter-mode tunneling that locally equilibrates the modes and drives the 
system to a RG-fixed point with the expected Hall conductance and minimal fractional charge $e^\star$~\cite{KaneFisherPolchinski,LevinHalperinRosenow,LeeRyuNayakFisher}. At the disorder-dominated phase the edge theories are 
$U_c (1)\times \overline{G}$ CFTs made of a chiral charged boson $\phi_\rho$ and counter-propagating neutral modes. The latter are described by $G=SU(2)_1$ and $G=SU(2)_2$ CFTs~\cite{CFT} for $\nu=2/3$ and the APF respectively.
The algebra $SU(2)_1$ is realized by a neutral chiral boson $\phi_n$~\cite{KaneFisherPolchinski}; $SU(2)_2$ can be written as Ising$\times U(1)$, with the $U(1)$ part realized by the neutral boson $\varphi_n$. The Ising part contains the fields $I$, $\psi$, $\sigma$ with scaling dimensions $0$, $\frac{1}{2}$, $\frac{1}{16}$ respectively~\cite{CFT,LevinHalperinRosenow,LeeRyuNayakFisher,Carrega}. 

Both $\nu=2/3$ and the APF at the disorder-dominated phases admit three quasiparticle fields with minimal conformal dimension and a multiplet of electron fields with the same total conformal dimensions $h$ (reported in Tab.~\ref{tableg})~\cite{CFT,LevinHalperinRosenow,LeeRyuNayakFisher,Ferraro,Carrega}. 
We shall use the notation $\Psi_{(\ell,m)}^{qp/e}$ to denote the quasiparticle/electron fields with $m$-independent conformal dimension ($\ell$ and $m$ denote the quantum numbers  resulting from the SU$(2)$ symmetry of the neutral sector, which is an emergent property of the edge theory near the disorder-dominated RG fixed point~\cite{KaneFisherPolchinski,KaneFisher,
LevinHalperinRosenow,LeeRyuNayakFisher}).
For the $\nu=2/3$ state the quasiparticle fields are $\qp_{(1,\pm)}=e^{\pm i\phi_n/\sqrt{2}}e^{i\phi_\rho/\sqrt{6}}$ and  $\qp_{(0,0)}=e^{i2\phi_\rho/\sqrt{6}}$; the doublet of electron fields is written as 
$\ef_{(1,\pm)}=e^{\pm i\phi_n/\sqrt{2}}e^{i\sqrt{3/2}\phi_\rho}$. The APF quasiparticle fields are $\qp_{(1,\pm)}=\sigma e^{\pm i\varphi_n/\sqrt{4}}e^{i\phi_\rho/\sqrt{8}}$ and 
$\qp_{(0,0)}=e^{i\phi_\rho/\sqrt{2}}$ and a triplet of electron fields is present: $\ef_{(2,\pm)}=e^{\pm i\varphi_n}e^{i\sqrt{2}\phi_\rho}$ and $\ef_{(2,0)}=\psi e^{i\sqrt{2}\phi_\rho}$.

The energy of the edge state depends on the quantized charge $Qe=N e+q e^\star$ present on the edge, on the variation of the area of the fluid $S-S_0$ with respect to its value at the center of the plateau ($Q=N=0$, $S=S_0$), and on the neutral index $\ell$ through the neutral conformal dimension $h_\ell$.
The ground state energy is~\cite{IlanGrosfeldStern,IlanGrosfeldSchoutensStern,CappelliViola}
\begin{eqnarray}\label{eq:Etot}
E(Q,S,B,\ell) = \frac{ v_c}{2 R \overline{\nu}}\left(Q-\frac{\overline{\nu}B(S-S_0)}{\phi_0}\right)^2+\frac{ v_nh_\ell}{R},
\end{eqnarray}
with $\frac{v_{c(n)}}{R}$ the energy scales of the charge (neutral) modes, $\overline{\nu}$ the fractional part of $\nu$, and $\phi_0$ the flux quantum.

\begin{table}
\begin{tabular}{c|c|c|c|c|c|c|c|}
\cline{2-8}&\multicolumn{3}{|c|}{$\nu=$2/3}&\multicolumn{4}{|c|}{APF}\\
\cline{2-8}& $\qp_{(1,\pm)}$& $\qp_{(0,0)}$&$\ef_{(1,\pm)}$&$\qp_{(1,\pm)}$&  $\qp_{(0,0)}$&$\ef_{(2,\pm)}$& $\ef_{(2,0)}$\\\hline
\multicolumn{1}{|c|}{$h$}&$1/3$&$1/3$&$1$&$1/4$&$1/4$&$3/2$&$3/2$\\
\multicolumn{1}{|c|}{$h_n$}&$1/4$&0&$1/4$&$3/16$&$0$&$1/2$&$1/2$\\
\multicolumn{1}{|c|}{$q$}&$1$&$2$&$3$&$1$&$2$&$4$&$4$\\
\hline
\end{tabular}\\
\caption{The conformal data for the quasiparticles' and electrons' fields: total and neutral scaling dimensions $h=h_c+h_n$ and $h_n\equiv h_\ell$, $q$ is the
charge in unit of $e^\star$. 
}\label{tableg}
\end{table}

{\it{Calculation of degeneracies}:---}Due to the presence of the neutral modes, the bulk state and the number of electrons on the edge do not always fully determine the ground state of the edge. This is manifested by the presence of several distinct electron fields that can tunnel into the dot at the point of lifting of the Coulomb blockade~\cite{Georgiev,CappelliZemba,CappelliViola,IlanGrosfeldStern,IlanGrosfeldSchoutensStern}, and as we see in the next section, this will affect the shape of the Coulomb blockade peaks. Here we find the number of ground states as function of the number of electrons on the edge and bulk state.
Formally this pattern of degeneracies can be extracted from the edge partition function~\cite{Georgiev,CappelliZemba,CappelliViola} or directly by fusion of the operators~\cite{IlanGrosfeldStern,IlanGrosfeldSchoutensStern}.  In the following we will resort to the latter method. 
Necessary conditions are  $T\ll\frac{v_n}{R}$ and $\frac{v_n}{R}<\frac{v_c}{R}$~
\cite{IlanGrosfeldStern,IlanGrosfeldSchoutensStern,CappelliViola}.
The p-h breaking is pronounced for $T\ll\frac{v_c}{R}$, while the constraint $\frac{v_n}{R}<\frac{v_c}{R}$ guarantees that tunneling will involve the neutral mode, and hence be sensitive to the pattern of degeneracies.

Consider the case of an APF droplet with an initial even number $N$ of electrons and no quasiparticles in the bulk. In this state the edge is in the trivial ground state described by the unique 
identity operator $I$ in both the charge and neutral sectors~\cite{IlanGrosfeldStern,IlanGrosfeldSchoutensStern}. The incoming electrons carry a topological charge of $(q=4,\ell=2)$. While the topological charge $q$ is  additive, the topological charge $\ell$ obeys the SU$(2)_2$ fusion rules, for which $2\times \ell=2-\ell$. We now consider changing the area of the dot $S$ by means of a side gate until the system reaches the first intersection $E(Q,S,B,\ell)=E(Q+1,S,B,2-\ell)$ marking a Coulomb peak (the distance between peaks has been already evaluated in~\cite{SternHalperin,IlanGrosfeldStern,IlanGrosfeldSchoutensStern,SternRosenowIlanHalperin,BondersonNayakShtengeldo,CappelliViola}). Then, it becomes energetically favorable to tunnel in one of three flavors of electrons $\ef_{(2,m)}$, say $\ef_{(2,0)}$, which consequently gets trapped in the dot. This means that at this first intersection, the degeneracy is $3$. Next, we further vary the size of the dot until the next intersection is reached. Now the energy is minimized only if the electron entering the dot is the one described by the field $\ef_{(2,0)}$ that brings the neutral edge back to the trivial sector
(similarly, $\ef_{(2,\pm)}$ is necessarily followed by $\ef_{(2,\mp)}$). Consequently, there
is no degeneracy in the choice of the electron fields. The pattern of the degeneracies as function of the electron number for this case is $1,3,1,3,\ldots$. 
 For $\nu=2/3$, in the absence of quasiparticles in the bulk, the edge is in the unique identity sector $I$. At the first intersection each of the two electrons $\ef_{(1,\pm)}$ may enter the dot and the degeneracy is $2$. Changing $S$ further a second resonance is reached and the energy in Eq. \eqref{eq:Etot} is minimized iff the incoming electron has the opposite neutral charge ($\ef_{(1,\mp)}$); hence the degeneracy is $1$. A similar analysis can be straightforwardly extended (see supplementary material) to the other cases and we obtain the following patterns of degeneracies: 
\begin{eqnarray}
  \label{eq:deg23} \nu=2/3 \quad\quad   && 1\mapsto 2\mapsto 1\mapsto 2\cdots\\
  \label{eq:degAPFeven} \mbox{APF-even} \quad\quad && 1\mapsto 3\mapsto 1\mapsto 3\cdots\\
  \label{eq:degAPFodd}\mbox{APF-odd} \quad\quad && 2\mapsto 2\mapsto 2\mapsto 2\cdots
\end{eqnarray}
Here APF-even and APF-odd denote the APF with an even and odd number of quasi-particles in the bulk. The presence of a finite temperature can drive the system slightly away from this RG fixed point providing a finite lifetime to the energy levels at the edge of the dot (as energy transfer between the neutral and charge modes becomes possible)~\cite{KaneFisherPolchinski,KaneFisher}. At the absence of any intermediate energy scales, and as long as $T\ll v_n/R$, the effects of degeneracy are robust.

{\it{Thermoelectric response of a dot}:---}
In order to analyze the response of the dot, let us first consider two terminal transport across a double QPC geometry embedded in a $\nu=5/2$ or $\nu=2/3$ quantum Hall liquid (see Fig.~\ref{fig:setup}). 
For $\nu=5/2$, we assume that the gate voltages at the QPCs are tuned such that only the higher $\overline{\nu}=1/2$ edge is strongly 
backscattered defining a QD with circulating $\overline{\nu}=1/2$ edge states.
To have a concrete starting point for the transport calculation we assume that the magnetic field is tuned initially to the value at the center of the plateau and the number of electrons in the dot is divisible by $2$. Hence, no quasiparticles are trapped in the bulk of the dot and the charge at the edge 
 has been set to zero corresponding to setting the Fermi level at the dot's edge state to zero.

The tunneling coupling $\Gamma_{L/R}$ between the dot and the L-R edges (see Fig.~\ref{fig:setup}) are taken to be weak $\Gamma_{L/R}\ll  T\ll v_{c,n}/R$ moreover electrical and thermal bias are small compared to the typical energy scale associated with electron states in the dot, \emph{i.e.} $e V, \Delta T\ll v_{c,n}/R $.  Therefore, we can safely ignore contributions to the current coming from other electron levels except for the levels at resonance.
 Assuming a dot of radius $1 \mu$m and $v_n\simeq 4\times 10^3$m/s~\cite{WanHu} the constraint $T\ll v_n/R$ restricts the electron bath temperature to be $T\ll 30$ mK.
Electrons can tunnel into the dot iff the side gate is tuned such that $E(Q,S,B)=E(Q+1,S,B)$~\cite{IlanGrosfeldStern,IlanGrosfeldSchoutensStern} (neutral indices suppressed) and the Fermi level in the lead is in tune with the dot Fermi level~\cite{Beenakker1991,BeenakkerStaring}.
Hence calculating transport across the system is reduced to calculating current through a single resonant level in the sequential tunneling regime~\cite{Beenakker1991,BeenakkerStaring,Furusaki}. The state of the dot driven out of equilibrium can be discussed in terms of occupancy of the level, while transport across the system is expressed in terms of rate equations~\cite{Beenakker1991,BeenakkerStaring}.

As discussed earlier, the degeneracy ${\cal D}_a$ of the electron addition spectrum depends on the parity of the electrons number $N$ in the dot
for both Hall states considered here (see Eqs.~(\ref{eq:deg23},\ref{eq:degAPFeven},\ref{eq:degAPFodd})).
 However, multiple occupancies of degenerate levels are forbidden by charging energy~\cite{Beenakker1991,IlanGrosfeldStern}. Hence the states of the level which participates in transport are the empty and the singly occupied states. 
For the general case of transitions between a ${\cal D}_{a}$-fold to ${\cal D}_{a+1}$-fold degenerate levels, with $N$ and $N+1$ electrons in the dot respectively [initially all the levels are assumed to have the same tunnel rates $\Gamma^{\pm}$ to tunnel in (out)
], the rate equation is given by
\begin{eqnarray}
\dot{P}_0&&=-{\cal D}_{a+1}\,\Gamma^+ P_0+{\cal D}_{a}\,\Gamma^- P_1 \nonumber\\
\dot{P}_1&&=-{\cal D}_{a}\,\Gamma^- P_1+{\cal D}_{a+1}\, \Gamma^+  P_0,
\label{rate}
\end{eqnarray}
where $P_0$ is the probability of the dot's level to be empty 
while $P_1$ is the probability that a single electron occupies the level, such that $P_0+P_1=1$. 
Eqs.~\ref{rate} can be solved in the steady state limit to obtain $P_0={\cal D}_{a} \Gamma^{-}/({\cal D}_{a}\Gamma^{-}+{\cal D}_{a+1} \,\Gamma^{+})$ and $P_1={\cal D}_{a+1} \,\Gamma^{+}/({\cal D}_{a}\Gamma^{-}+ {\cal D}_{a+1} \,\Gamma^{+})$. 
The effective rates $\Gamma^{\pm}$ are sums of rate due to tunneling between the level and the left 
$\Gamma^{\pm}_L$ and right edge $\Gamma^{\pm}_R$: $\Gamma^{\pm}=\Gamma^{\pm}_L + \Gamma^{\pm}_R$. Hence the 
tunneling current flowing from the left (right) edge into the dot is simply given by 
$I_{R/L}=\pm e/(2\pi ) {\dot P}_{0,{R/L}} $ where $\dot P_{0,{R/L}}$ can be extracted from R.H.S. of the first equation in Eq.~\ref{rate} simply by
first substituting $\Gamma^{\pm}=\Gamma^{\pm}_L + \Gamma^{\pm}_R$ and then collecting terms corresponding to 
$ \Gamma^{\pm}_L$ or $ \Gamma^{\pm}_R$. For algebraic convenience we choose a symmetric definition of current given by $I=(I_L+I_R)/2=e/(4 \pi ) ~ ({\dot P}_{0,R}-{\dot P}_{0,L})$.

Following Furusaki~\cite{Furusaki} the tunnel rates evaluated to lowest order in tunneling amplitude $\Gamma_{L/R}$ are given by 
\begin{eqnarray}\label{}
\Gamma^{\pm} &=& T_R^{} \, e^{\mp \frac{\varepsilon}{2T_R}} \, \gamma_{R} +
T_L^{} \, e^{\mp \frac{\varepsilon - e V}{2T_L}} \, \gamma_{L}~, \nonumber\\
\gamma_{f} &=& \frac{\Gamma_{f}^{}}{2 \pi T_{f}^{}}
 \left(\frac{\pi T_{f}}{\Lambda}\right)^{2 h-1}~ \frac{\left\vert\Gamma \left(h + i \frac{\varepsilon - eV\delta_{f,L} }{2\pi T_f}\right)\right\vert^2}{\Gamma(2 h)}
\end{eqnarray}
where $f=L,R$, $T_L=T+\Delta T$ and $T_R=T$ are the temperatures of the left and the right edge respectively and  $\varepsilon=E(N+1,S,B)-E(N,S,B)$. The linear response current can be obtained by expanding the above expressions 
to leading order in  $e V$ and $\Delta T$  as 
\begin{eqnarray}
  I &=& \frac{e\Gamma_0 {\cal D}_{a+1}}{(2\pi)^2}  \frac{(\pi T/ \Lambda)^{2 h-1}}{\Gamma(2h)}  ~\left| \Gamma 
  \left(  h +i \frac{\varepsilon}{2\pi T}\right)\right| ^2 \left(P_0~e^{-\frac{\varepsilon}{2 T}}\right)  \nonumber \\
  && \left(\frac{\varepsilon}{2 T} \frac{\Delta T}{T} +\frac{eV}{2T}\right)
\equiv G_T \Delta T+ G e V.
\label{current2}
\end{eqnarray} 
where $P_0= {\cal D}_{a}/({\cal D}_{a}+{\cal D}_{a+1} e^{-\varepsilon/T})$ is the equilibrium occupancy of the level, $\Lambda$ is an energy cutoff, $\Gamma(x)$ is the Gamma function and assuming $\Gamma_L=\Gamma_R\equiv\Gamma_0$. 
As direct consequence of the sequential tunneling regime, the occupancies $P_{0,1}$ are independent of whether the edge state is a Fermi-liquid ($h=1/2$) or non-Fermi liquid ($h \neq 1/2$). 
  The ratio between the maximum of thermoelectric $G_T$ and the electric $G$ coefficients at the peaks (see Fig.~\ref{fig:patterns}), is  
$G_T^{max}/G^{max}= K^{\nu}_{{\cal D}_{a},{\cal D}_{a+1}}$. 
For $\nu=2/3$ we find $K^{2/3}_{2,1}= 0.85$, $K^{2/3}_{1,2}=1.42$ and for APF 
$K^{APF}_{3,1}=0.76,~K^{APF}_{2,2}=1.18,~K^{APF}_{1,3}= 1.76$.
The absolute value of the ratio of height of the two lobes of the thermoelectric coefficient peak constitutes an alternative measure of the degeneracies ${\cal D}_a$. 
For $\nu=2/3$ the ratio is 1.66 or its inverse depending on the parity of $N$ (with $({\cal D}_a,{\cal D}_{a+1})=(1,2),~(2,1)$). For APF in absence of quasiparticles in the dot the ratio is 2.31 or its inverse for consecutive peaks 
($({\cal D}_a,{\cal D}_{a+1})=(1,3),~(3,1)$)
and is one otherwise (${\cal D}_{a+1}={\cal D}_{a}=2$). 
The conclusions of Eq.~\eqref{current2} remain intact also in the presence of more general tunneling processes that allow mixing of electrons with different neutral number~\cite{FieteBishara} and different couplings to the leads as long as the tunneling matrices to the two leads are proportional (see supplementary material for details.)
An alternative edge theory for $\nu=5/2$ is given by the edge-reconstructed PF (RPF)~\cite{OverboschWen}, in which an upstream neutral mode exists. Since the  thermal Hall conductivity for RPF (APF) is $3/2$  ($-1/2$) [$\pi k_b^2T/6$], it is expected that the decay of the thermal current is faster (slower) with respect to $\nu=2/3$ ~\cite{kanefisherIQ}, leading to a different functional dependence for $\Delta T$ on the applied $V$. Moreover, the APF and RPF edge theories are topologically distinct~\cite{LevinHalperinRosenow,LeeRyuNayakFisher,OverboschWen} and their associated patterns of degeneracies are different.

{{\it Discussion:}---}For a system of two electron reservoirs coupled through a QD, the degeneracy of a resonant energy level in the dot manifests itself in two ways: first, it appears as a scale factor in the 
total current representing the number of independent channels for transport; and, second, via its appearance in the expression for the dot's occupancy. The latter has a non-trivial effect on electric and thermoelectric transport as we now describe. 

We first consider electric response ($V\neq 0$, $\Delta T=0$).
 For ${\cal D}_a={\cal D}_{a+1}$ the center of the peak in the linear electric coefficient $G$ is at $\varepsilon=0$. 
However, when ${\cal D}_a\neq{\cal D}_{a+1}$, the peak is shifted to a finite value of $\varepsilon$ which depends on both ${\cal D}_a$, ${\cal D}_{a+1}$ and $T$~\cite{Beenakker1991,Georgiev}. 
 This directly relates to 
the equilibrium probability for zero occupancy $P_0$: for ${\cal D}_a={\cal D}_{a+1}$, it is exactly $1/2$ at resonance ($\varepsilon=0$) which leads to a maximum in current.  
However, when the system crosses between states with different degeneracies, detailed balance requires that $P_0(\varepsilon=0)\neq 1/2$ hence the peak is shifted to a finite value of $\varepsilon$.
These considerations would lead to a shift of $(T\ln{{\cal D}})/2$ (\emph{i.e.} half of the entropy associated with the degeneracy of the level) for the peak between states with degeneracy $1$ and ${\cal D}$ for the simplified case of free 
electrons leads ($h=1/2$).
 For our case of non-Fermi liquid leads, $h$-dependent terms in Eq.~\eqref{current2} will modify the formula for the shift.  

More pronounced consequences of the degeneracy occur for the thermoelectric coefficient $G_T$ ($V=0$, $\Delta T\neq 0$). 
The thermoelectric coefficient has an odd 
profile as long as $P_0$ respects the p-h  
 symmetry around the Fermi level of the lead with $P_0=1/2$ for $\varepsilon=0$.
This symmetry in $P_0$ is lifted as we switch on a non-constant degeneracy and this results in an asymmetry between hole and particle currents, see Fig.~\ref{fig:patterns}. The presence of an asymmetry can serve as a direct measure of the presence of such degeneracies, and gives a local probe for the value of the degeneracy (while a measurement of the shift discussed in the previous paragraph requires, for example, to measure the distance between two peaks at two different temperatures, which is expected to be more difficult). This is one of the central points of our letter. 
 
As discussed earlier both $\nu=2/3$ and APF states have degeneracies in their spectrum for the addition of an extra electron to the dot 
as given by Eqs.~(\ref{eq:deg23},\ref{eq:degAPFeven},\ref{eq:degAPFodd}). The degeneracies depend on whether $N$ is even or odd and, in the APF case, on the bulk charge. Hence in conclusion 
we predict that for both of these states the thermoelectric peaks will show strong disparity due to successive appearance and disappearance of the degeneracy, Fig.~\ref{fig:patterns}, but for the APF there are two distinct patterns depending on the number of quasi-particles in the bulk. 
Note that the experiments reported in Refs.~\cite{BidOfek,DolevGross} confirmed that both $\nu=2/3$ and the APF have counter propagating charge and neutral modes using charge noise measurement in a zero current situation. On the contrary our proposal is expected to produce a current signal via 
thermoelectric effect hence circumventing the need for a technically difficult noise measurement. Additionally it also provides 
unique direct information regarding the edge spectrum via degeneracy induced features in the thermoelectric current.

{{\it Note}:---After this manuscript has been posted on the web, an experiment carried out by I. Gurman, R. Sabo et al (Ref~\cite{GurmanSabo}) was reported, in which our predicted thermoelectric effect was observed. 

{{\it Acknowledgments}:---We thank A. Bid, G. Campagnano, A. Cappelli, M. Dolev, D. Ferraro,  Y. Gross, I. Gurman, M. Heiblum, R. Sabo, A. Yacoby and in particular Y. Gefen, A. Kamenev and B. Rosenow for useful discussions.
SD thanks the Weizmann Institute of Science, Israel for their kind hospitality during his visit in July, December 2011. 
EG thanks the Israel Science Foundation (grant no.~401/12) and the European Union's Seventh Framework Programme (FP7/2007-2013) under grant agreement no.~303742, and the Aspen Center for Physics and Microsoft's Station Q for their hospitality.
 AS thanks the US-Israel Binational Science Foundation, the Minerva foundation and Microsoft's Station Q for financial support. 
GV thanks Feinberg and A. D. Riccia foundations for financial support.
}

\bibliographystyle{apsrev4}
\bibliography{thermoelectric}

\renewcommand{\thefigure}{S\arabic{figure}}
 \setcounter{figure}{0}
\renewcommand{\theequation}{S.\arabic{equation}}
 \setcounter{equation}{0}
 \renewcommand{\thesection}{S.\Roman{section}}
\setcounter{section}{0}

\section*{Thermoelectric probe for neutral edge modes in the fractional quantum Hall regime\\ supporting material}

\section{Additional details of the theory for the edge states}

 In this section we summarize the details of the edge theories used to derive the results in the main text.

 The charge part of the edge theory is described by a chiral boson $\phi_\rho$  that realizes a $U_c (1)$ CFT (chiral Luttinger liquid).
 Normalization of $\phi_\rho$ is such that the electronic charge density is given by 
$\rho_c=\frac{\sqrt{\overline{\nu}}}{2\pi}\partial_x\phi_\rho$ (with $\overline{\nu}$ the fractional part of $\nu$)~\cite{TQC}. 
The presence of a charge $q e^\star$ ($q$ an integer) in the bulk affects $\phi_\rho$ via the boundary condition 
$\sqrt{\overline{\nu}}(\phi_\rho(2\pi R)-\phi_\rho(0))=2\pi q e^\star/e$;
here  $2\pi R$ is the length of the edges. 
Accordingly, the charge $Q e$ assumes only discrete values $Qe=N e+q e^\star$
 equal to the charge added to the edge of the fluid with respect to the value at the center of the plateau, $Q=N=0$.
If the area of the Hall fluid is shifted away from $S_0$, its value at the center of the plateau, by $S-S_0$, the charge Hamiltonian becomes
$
{\hat H}^\nu_c=\frac{v_c}{4\pi}\int\mathrm{d}x\left(\partial_x\phi_\rho-\frac{\sqrt{\overline{\nu}}B(S-S_0)}{R\phi_0}\right)^2 
$
~\cite{IlanGrosfeldStern,IlanGrosfeldSchoutensStern},
with $v_c$ the velocity of $\phi_\rho$. The ground state energy for an edge state with charge $Q$ corresponds to uniform density and is 
$E_c(Q,S,B)=\frac{ v_c}{2 R \overline{\nu}}\left(Q-\frac{\overline{\nu}B(S-S_0)}{\phi_0}\right)^2$.

Neutral modes introduce additional contributions to the energy~\cite{TQC,IlanGrosfeldStern,IlanGrosfeldSchoutensStern,CappelliViola}. 
The $\nu=2/3$ state supports a neutral chiral boson $\phi_n$ subject to the quantization condition $\sqrt{2}(\phi_n\small{(2 \pi R)}-\phi_n\small{(0)})=2\pi \mathbb{Z}$~\cite{KaneFisherPolchinski}. The APF's neutral edge theory can be written as Ising$\times U(1) $ whose boson $\varphi_n$ is subject to the constraint $\varphi_n\small{(2\pi R)}-\varphi_n\small{(0)}=\pi \mathbb{Z}$~\cite{LevinHalperinRosenow,LeeRyuNayakFisher,Carrega}.
In general, the Hamiltonian of the edge modes is proportional to the Casimir of the corresponding Virasoro algebra $L_0$~\cite{CFT}: 
${\hat H}^\nu_n=\frac{ v_n}{ R}L_0$. The eigenvalues of ${\hat H}^\nu_n$ are $E_n=\frac{v_n}{ R}(h_n+{\kappa})$, $h_n$ being the conformal dimensions of the primary fields~\cite{CFT} reported in Tab. I in the main text and the positive integers ${\kappa}$ label excited states. Apart from the explicit dependence on $R$, the neutral modes are assumed to be unaffected by variations of the dot area.

The total energy of the combined edge state depends
on the charge Q and on the neutral index $\ell$. The
ground state energy  assumes the form of Eq. 1 of the main text.

\section{Calculation of degeneracies}

Here, we derive the degeneracies of the lowest Hall dot levels ${\cal D}_a$ for consecutive tunneling of electrons for both $\nu=2/3$ and the $\nu=5/2$ Anti-Pfaffian (APF) states. 
The degeneracies will be obtained from the grand-canonical partition function of the edge (Z)~\cite{CappelliZemba,Georgiev,CappelliViola} and also via the fusion of operators~\cite{TQC,IlanGrosfeldStern,IlanGrosfeldSchoutensStern}. 
The relevance of the degeneracies in Hall quantum dot (QD) physics was first pointed out by L. S. Georgiev~\cite{Georgiev}. 

For completeness, we report the essential details of the conformal field theory (CFT)~\cite{CFT} description for the bulk and edge of quantum Hall fluids~\cite{TQC,MooreRead}.
The Laughlin states at filling factors $\nu=1/p$ are described by a single charge chiral $U(1)$ CFT with $p$ primary fields $e^{iq\phi_\rho/\sqrt{p}}$, $q=0,\cdots, p-1$ and $h=q^2/(2p)$.
 Other non-Laughlin states, including $\nu=2/3$ and the APF, have more elaborate effective descriptions composed of charge and neutral algebras.
Moreover, electron fields have non-trivial neutral part that modifies the boundary condition (as reported in the main text) of $\phi_\rho$.
For $\nu=2/3$ (APF), such condition admits $p=6$ ($p=8$) distinct primary fields with charges $qe^\star$ and $q=0,\cdots, 5~(7)$. 
The Abelian neutral algebra for $\nu=2/3$, $\overline{SU(2)}_1$, admits two primary fields $I=\psi_{(\ell=0)}=e^{i0\phi_n}$ and $\psi_{(\ell=1)}=e^{\pm i\phi_n/\sqrt{2}}$, with the latter constituting two distinct primary fields $\psi_{(\ell=1,\pm)}$ under the Virasoro Algebra obtained from $\overline{SU(2)}_1$.
The overline indicates that the chirality of neutral modes is opposite with respect to the charge ones.
The non-Abelian $\overline{SU(2)}_2$ algebra for APF admits three primary fields, with conformal dimensions $(h_n)_\ell=0,3/16,1/2$ for $\ell=0,1,2$ respectively.  We utilize the representation of $SU(2)_2=$Ising$\times U(1)$~\cite{CFT,Carrega} where
\begin{align}\label{sup-eq:phi}
\psi_{(\ell=0,0)}&=I~,&&\psi_{(\ell=1,\pm)}=\sigma e^{\pm i\varphi_n/2}~, 
\\ \psi_{(\ell=2,\pm)}&=e^{\pm i\varphi_n}~,&&\psi_{(\ell=2,0)}=\psi~,
\end{align}
where $\sigma$ and $\psi$ are the spin and the Majorana fields of the the Ising CFT respectively.
In general, $\ell$ labels also the representation of the algebra $\psi_{(\ell)}$.
The fusion rule $\psi_{(\ell)}\times\psi_{(\ell=2)}=\psi_{(2-\ell)}$ of $\overline{SU(2)}_2$
 is applicable here.
The quasiparticles of the fluids are described by the product of primary fields in charge and neutral algebra $\Psi_{(\ell,m)}=\psi_{(\ell,m)}e^{iq\phi_\rho/\sqrt{p}}$, respecting the condition that $[\ell-q]_2=0$ (\emph{i.e.} $\ell\equiv q\,$mod$\,2$)~\cite{TQC}.

We are going to introduce partition functions Z for droplets of Hall fluids~\cite{CFT,CappelliZemba,CappelliViola}. The partition functions Z have been obtained by keeping a single chirality of the corresponding partition functions on the annulus geometry, where they respect modular invariance~\cite{CFT}. The energy of the edge states can be expressed through the Casimir operators of the algebras $\hat{H}^{\nu}_c=v_c L_0^{c}/R$ and $\hat{H}^{\nu}_n=v_nL_0^n/R$. 
The partition function is defined as ${\mathrm Z}(\beta,\mu)=\mathrm{Tr} e^{-\beta (\hat{H}^{\nu}_c+\hat{H}^{\nu}_n)+\beta\mu\hat Q}$, where the trace is over the edge states respecting the boundary conditions fixed by the bulk quasi-particles.
The possible boundary conditions are in correspondence with the edge topological sectors (or topological charges)~\cite{TQC,MooreRead}. Electron fields have trivial topological charge. Therefore, fields or edge states that can be obtained one from the other by fusing with electrons share the same topological charge: in this sense, the electron fields extend the chiral algebra~\cite{TQC,MooreRead,CappelliViola}. 
Hence, for a given sector the partition function Z will be given by trace over all ground states (primary fields operating on the vacuum) corresponding to the sector and all excited states (``descendant states'') over these ground states.
For the Jain states the topological charge $q$ is the number 
of fractional charges $-e^\star$ in the bulk modulo the electron charge, $q\,$Mod($e/e^\star$)~\cite{CappelliZemba}. 
In the APF state the topological charge is labeled by both the number of fractional charges $q$ and the topological number $\ell=0,1,2$ representing the non-Abelian charge in the droplets~\cite{TQC,IlanGrosfeldSchoutensStern,CappelliViola}. 
The expression of the partition function Z is expected to take the form
 ${\mathrm Z}(\beta,\mu)=\sum_s {\cal D}_s e^{-\beta E_s+\mu\beta Q_s}$, where $s$
labels all the possible energy and charge eigenvalues. Therefore, ${\cal D}_s$ is the number of available states for the  charge and energy labeled by $s$. 
The neutral and charge contributions to the partition function Z can be collected in the form of characters, denoted $K_q$ for the charge part and for the neutral part $\overline{\Theta}_\ell$ for $\overline{SU(2)}_1$ and $\overline{\chi}_\ell$ for $\overline{SU(2)}_2$ algebras~\cite{CappelliZemba,CappelliViola}. 
The neutral and charge characters (defined in Ref.~\cite{CFT}) are power series in $e^{-\beta v_{n,c}/R}$, respectively and are in one-to-one correspondence with the primary fields. 
The complete expression the partition functions Z for the $\nu=2/3$ and the APF states are 
\begin{align}\label{sup-eq:Z23}
{\mathrm Z}_q=&K_{q}(\beta,\mu)\overline{\Theta}_{[q]_2}(\beta)+K_{q+3}(\beta,\mu)\overline{\Theta}_{[q+1]_2}(\beta)~,\\
\label{sup-eq:ZAPF}
{\mathrm Z}^\ell_q=&K_{q}(\beta,\mu)\overline{\chi}_\ell(\beta)+K_{q+4}(\beta,\mu)\overline{\chi}_{2-\ell}(\beta)\,,
\end{align}
respectively, where ($p=2e/e^\ast$)~\cite{CappelliZemba}
\begin{eqnarray}\label{sup-eq:K}
K_{q}(\beta,\mu)=\sum_{m\in\mathbb{Z}}e^{-\beta E_c(\frac{2q}{p}+2m,S,B)+\mu\beta(\frac{2q}{p}+2m)}\,,
\end{eqnarray}
neglecting irrelevant coefficients at $T\ll v_c/R$. The factor proportional to $\mu\beta$ is the charge $Q=2q/p+2 m$ and the sum over $m$ in (\ref{sup-eq:K}) counts the addition of two electron charges. Hence, the first (second) term of R.H.S in Eqs. (\ref{sup-eq:Z23},\ref{sup-eq:ZAPF}) represents the contributions for even (odd) number of electrons $N$ added to the droplet's edge.
We consider  $T \ll\frac{v_n}{R}<\frac{v_c}{R}$, therefore only the lowest energy states of the dot are accessible. 
The low temperature expansion of the characters $\overline{\Theta}_\ell(\beta)$ and $\overline{\chi}_\ell(\beta)$ hence encodes the information we seek and is given by 
${\cal D}_\ell e^{-\frac{\beta v_n}{R}(h_\ell-c/24)}$ where $c$ is the central charge of the respective CFT~\cite{CFT}. 
The degeneracies assume the values ${\cal D}_\ell=1+\ell$ with $\ell=0,1$ for $\nu=2/3$ and $\ell=0,1,2$ for the $\nu=5/2$ APF. The charging energy, assumed to be the largest energy scale, fixes the charge degeneracy to be one.

The degeneracy patterns can be obtained from the partition function Z as follows~\cite{CappelliZemba,CappelliViola,Georgiev}. 
Consider the APF dot with a given bulk topological charge: the edge will be in sector $(q,\ell)$ such that the whole dot has trivial topological charge~\cite{IlanGrosfeldSchoutensStern}. 
For  $T\ll\frac{v_n}{R}<\frac{v_c}{R}$ the edge is in the state that minimizes the total energy $E$ in Eq.~(1) 
 as function of $S-S_0$ for the given $(q,\ell)$ of the edge.
 At $S-S_0\simeq 0$ the lowest energy state appears in $K_q\overline{\chi}_\ell$, that corresponds to $N=0$, and has degeneracy ${\cal D}_{\ell}$. Charge fluctuations are suppressed until, acting with the side gate voltage, the system reaches the resonant point $S_1$ \emph{s.t.} $E(q/4,S_1,B,\ell)=E(q/4+1,S_1,B,2-\ell)$. The tunneling of one electron is allowed, that gets trapped into the dot and the charging energy blocks further tunneling~\cite{IlanGrosfeldSchoutensStern}. 
 Now, the edge is described by the second term in Eq.~\eqref{sup-eq:ZAPF} $K_{q+4}\overline{\chi}_{2-\ell}$ with degeneracy ${\cal D}_{2-\ell}$. 
Further variation of the size of the dot permits to reach a new resonant point
 $E(q/4+1,S_2,B,2-\ell)=E(q/4+2,S_2,B,\ell)$ 
and the edge goes back into a state in $K_q\overline{\chi}_{\ell}$ (${\cal D}_{\ell}$), and so on. 
The patterns of peaks is ${\cal D}_\ell\mapsto{\cal D}_{2-\ell}\mapsto{\cal D}_\ell\cdots$ and $q$ independent; Eqs.~(3,4) are recovered for both even and odd number of quasiparticles in the bulk (\emph{i.e.} $\ell=0$ and $\ell=1$ respectively). 
 The distance, on $S$ axis, between the peaks is given by the equations of resonance as function of $S$~\cite{IlanGrosfeldSchoutensStern,CappelliViola} which we repeat for completeness. For even number of bulk quasiparticles the distance between peaks are alternately $\Delta S_1$ (with ${\cal D}=3$) and  $\Delta S_2$ (with ${\cal D}=1$), with
\begin{eqnarray}\label{sup-eq:DS}
\Delta S_i&=\frac{e}{n_0}\left(1+(-)^i~\overline{\nu}~\frac{v_n}{v_c}\right)\,.
\end{eqnarray} 
For odd number of bulk quasiparticles the peaks are all at distance $\Delta S=e/n_0$ and with degeneracy ${\cal D}=2$.

Next, we compute the same degeneracy patterns using the other method of the fusion of the operators. 
Let us first consider $\nu=2/3$ QD with a quasihole with charge $-e/3$ in the bulk and $S-S_0$ just big enough to accommodate such charge. The edge is therefore labeled by $q=1$ (with  $Q=e/3$ on it).
 The most relevant term in the partition function Z at low $T$ appears in the first term of (\ref{sup-eq:Z23}) $K_{1}\overline{\Theta}_1$ whose low temperature expansion gives ${\cal D}_{1}=2$. This is confirmed in the operator picture: the edge's sector can be in both $\qp_{(1,\pm)}$ (suppose $\qp_{(1,+)}$) Virasoro representations with the same energy $E(1/3,S,B,1)=E_c(1/3,S,B)+\frac{v_n}{4R}$. Varying $S$, we get a resonant point and one of the two electron fields $\Psi^e_{(1,\pm)}$ (defined in the main text) can enter into the dot. 
 From the fusion of the fields 
$\qp_{(1,+)}\cdot \ef_{(1,-)}=e^{i4\phi_\rho/\sqrt{6}}$
  and
 $\qp_{(1,+)}\cdot \Psi^e_{(1,+)}=e^{i\sqrt{2}\phi_n} e^{i4\phi_\rho/\sqrt{6}}$ we see that the minimum energy state is obtained only if the entering electron has opposite neutral charge with respect to the edge state, \emph{i.e.}  $\ef_{(1,-)}$.
 The resonance is obtained at $E(1/3,S_1,B,1)=E_c(1/3+1,S_1,B)$ and the dot is now described by the lowest energy state in $K_{1+3}\overline{\Theta}_0$ for which ${\cal D}_{0}=1$. Following the tunneling, the neutral edge charge is trivial. The next resonance, $S_2$, for which $E_c(1/3+1,S_2,B)=E(1/3+2,S_2,B,\pm 1)$, can receive equivalently both electrons: the second peak has ${\cal D}_{1}=2$. 
Indeed, $e^{i4\phi_\rho/\sqrt{6}}\cdot\Psi^e_{(1,\pm)}$ have the same energy.
The process can be repeated iteratively to obtain the pattern in Eq. (2) of the main text.
Also for $\nu=2/3$ there exist a modulation of the peaks' distance: the peak with degeneracy ${\cal D}_{1}=2$ is followed by one at distance $\Delta S_1$ with ${\cal D}_{0}=1$, the third one will be at $\Delta S_2$  with ${\cal D}_{1}=2$ and so on. For $\nu=2/3$, $v_n$ in \eqref{sup-eq:DS} has to be substituted with $v_n/2$.
%

 The RPF state~\cite{OverboschWen} is an alternative candidate to describe the plateau at $\nu=5/2$. 
It is a derivative of the PF state in the presence of edge-reconstruction effects and admits upstream modes~\cite{OverboschWen}.
In the absence of interactions between counter-propagating edges the theory is described by an independent PF edge (going downstream) with additional $\nu=1$ modes propagating in opposite directions, each admitting trivial degeneracy patterns~\cite{CappelliViola} (the PF state, like all Read-Rezayi states~\cite{ReadRezayi-S}, has ${\cal D}=1$ for all peaks independently of the number of quasiparticles in the bulk)~\cite{CappelliViola}. This leads to a degeneracy pattern for the RPF which is independent of the presence of quasiparticles in the bulk. When interactions between edge states are switched on, the system may go into the Majorana-gapped phase~\cite{OverboschWen}, for which the third electron is gapped~\cite{OverboschWen} 
(and therefore degeneracy ${\cal D}=3$ can not be reached). This also results in a pattern of degeneracies that is different from Eqs. (3) and (4) in the main text.

The analysis presented above can be straightforwardly extended to other states of interest, \emph{i.e.} states for which upstream neutral modes and non-trivial sequences of degeneracies are expected, including the particle-hole conjugate of the Jain series (at $\nu=m/(2ms-1)$)~\cite{Frohlich,CappelliZemba}, anti-Read-Rezayi states~\cite{BisharaFieteNayak} and some of the Bonderson-Slingerland states~\cite{BondersonSlingerland-S}. This may be particularly relevant for detecting the properties of the observed $\nu=8/3$ plateau~\cite{DolevGross}.

\section{Thermoelectric response of a dot in presence of more general tunnel rates}

In the main text we solved the master equation for the steady state occupation of the dot for the case that the tunneling rates between the dot and the leads are equal and conserve the neutral charge (the leads and the dot are realized by separate quantum Hall droplets sharing the same state and filling factor $\nu$).  Here, we consider more general tunneling processes~\cite{FieteBishara} with mixing of electrons with different neutral number and demonstrate that their effect on transport through the dot is minor.
For concreteness, we discuss the case of $\nu=2/3$ close to the resonance between states with degeneracies $1$ and $2$ respectively. 

The tunneling Hamiltonian between the left/right ($f=L/R$) leads and the dot can be written as 
\begin{align}
H^{f}=&k_{f} \left(\begin{smallmatrix}(\Psi^e)^\dagger_{(1,+)},&(\Psi^e)^\dagger_{(1,-)}\end{smallmatrix}\right)_{f}
\left(
\begin{smallmatrix}
t_{11}& t_{12} \\ t_{21}&t_{22}
\end{smallmatrix}\right)
\left(\begin{smallmatrix}\Psi^e_{(1,+)}\\ \Psi^e_{(1,-)}\end{smallmatrix}\right)_{\mathrm{dot}}+\mathrm{h.c.}\label{sup-eq:tunh}
\end{align}
 where $k_{i}$ allows for different coupling strength for the left and right leads (we set $k_{R}=1$ and $k_{L}=k$).
The singular value decomposition guarantees that there are two unitary matrices $U$ and $V$ such that $t_{i,j}=(U \mathrm{Diag}({\tau_1,\tau_2})V^\dagger)_{i,j}$, hence Eq.~\eqref{sup-eq:tunh} can be rewritten as
\begin{align}\nonumber
H^{f}=k_{f} \left(\begin{smallmatrix}(\Psi^e)^\dagger_{1,s} ,&(\Psi^e)^\dagger_{1,a}\end{smallmatrix}\right)_{f} 
\left(
\begin{smallmatrix}
\tau_s &0\\ 0&\tau_a
\end{smallmatrix}\right)
\left(\begin{smallmatrix}\Psi^e_{1,s}\\ \Psi^e_{1,a}
\end{smallmatrix}\right)_{\mathrm{dot}}+\mathrm{h.c.}\,.
\end{align}
 where $\left(\begin{smallmatrix}\Psi^e_{1,s}\\ \Psi^e_{1,a}
\end{smallmatrix}\right)_{j}=M_J\left(\begin{smallmatrix}\Psi^e_{(1,+)}\\ \Psi^e_{(1,-)}\end{smallmatrix}\right)_{j}$ with $M_{\mathrm{dot}}=V$, $M_{f}=U$.
The case $\tau_{s/a}\neq 0$, which we now analyze, describes tunneling through a resonant two-fold degenerate level with different couplings for the two species of electrons (multiple occupancies of the degenerate level is forbidden).

Similarly to the main text, the system is described by a rate equation. 
The probability that the level will be empty is $P_0$, while the probability that the ``symmetric'' (``anti-symmetric'') state will be occupied is $P_s$ ($P_a$). The rate equations are
\begin{align}
\dot{P}_0&=-\left(W^{s0}+W^{a0}\right)P_0+W^{0s}P_s+W^{0a}P_a\nonumber\\
\dot{P}_{i=s,a}&=- W^{0i}P_i+W^{i0}P_0\label{sup-eq:rate}
\end{align}
 ($P_0+P_s+P_a=1$).
In the steady state limit ($\dot{P}_x=0$ for $x=0,s,a$) the solution of Eqs \eqref{sup-eq:rate} reads
\begin{align}
P_0&=\frac{W^{0a}W^{0s}}{W^{0a}W^{0s}+W^{0s}W^{a0}+W^{0a}W^{s0}}\nonumber\\
P_s&=\frac{W^{0a}W^{s0}}{W^{0a}W^{0s}+W^{0s}W^{a0}+W^{0a}W^{s0}}\nonumber\\
P_a&=\frac{W^{0s}W^{a0}}{W^{0a}W^{0s}+W^{0s}W^{a0}+W^{0a}W^{s0}}
\end{align}
The effective rates $W^{xy}$ (which are analoguous to $\Gamma^{\pm}$ in the main text) have contributions from the left and right leads $W^{xy}=W^{xy}_R+W^{xy}_L$. Using the same methods described in the main text we can define the current as 
\begin{align}\label{sup-eq:current}
I=\frac{e}{4\pi}\left[-(W^{s0}_R-W^{s0}_L+W^{a0}_R-W^{a0}_L)P_0\right.\nonumber\\
\left.+(W^{0s}_R-W^{0s}_L)P_s+(W^{0a}_R-W^{0a}_L)P_a\right]
\end{align}
We compute the current at linear order in $\Delta T$ and $V$.
Following Ref.~\cite{Furusaki} the rates are given by a straightforward generalization of Eq. (6) in the main text,
\begin{eqnarray}\label{sup-eq:defrates}
W^{0i}&=& T_R^{} \, e^{ \frac{\varepsilon}{2T_R}}\gamma_{R}^{0i}+
T_L^{} \, e^{ \frac{\varepsilon - e V}{2T_L}} \, \gamma_{L}^{0i}\\
W^{i0}&=& T_R^{} \, e^{ -\frac{\varepsilon}{2T_R}}\gamma_{R}^{i0}+
T_L^{} \, e^{ -\frac{\varepsilon - e V}{2T_L}} \, \gamma_{L}^{i0}
\end{eqnarray}
The functions $\gamma_{f}^{xy}$ are defined in the second line of Eq.~(6) in the main text with the constants $\Gamma_{f}^{xy}=\Gamma_{f}^{yx}$ (generalizing $\Gamma_{f}$ in the main text, $\Gamma_{f}^{0s/a}\propto \tau_{s/a}$) now depending on both the leads $f$ and the channel of the dot. 
For the tunneling Hamiltonian considered here we have in equilibrium $\Gamma_{L}^{xy}=\vert k\vert^2\Gamma_{R}^{xy}$, therefore $W^{xy}_L=\vert k\vert^2 W^{xy}_R=\frac{\vert k\vert^2}{1+\vert k\vert^2}W^{xy}$. 
        Hence we see that with a more general tunneling matrix at the two barriers, at equilibrium, $P_0=\frac{1}{1+2 \exp[-\epsilon/T_R]}$ which coincides with the result in the main text.
    Under similar conditions, for the case of ${\cal D}$-fold degenerate level tunnel-coupled to ${\cal D}$ channels in the lead with a general tunnel-matrix, 
    $P_0$ in equilibrium is given by $P_0=\frac{1}{1+ {\cal D}\exp[-\epsilon/T_R]}$  where all the eigenvalues of the tunneling-matrix have non-vanishing values. This implies that the effect of degeneracy to lowest order in tunneling 
in the linear response limit is independent of the details of the tunneling matrix as long as all the eigenvalues of the tunnel-matrix are non-zero. 
 The expansion at linear order in $V$ and $\Delta T$ of Eq.~\eqref{sup-eq:current} generates three type of terms given by the expansion of (i)/(ii) the denominator/numerator of the rates $P_{x=0,s,a}$; and (iii) the expansion of the difference of rates $W_R-W_L$. After some algebra one can deduce that the sum of type (i) terms vanishes. Type (ii) terms contribute a term which is proportional to $1- \vert k\vert^2$ hence vanishes for a symmetric dot. The final result is
\begin{eqnarray}
  I &=&\left(1+\frac{1-\vert k\vert^2}{2+2\vert k\vert^2}\right) \frac{e \vert k\vert ^2 (\Gamma_R^{s0}+\Gamma_R^{a0})}{(2\pi)^2\Gamma(2h) } \left( \frac{\pi T}{ \Lambda}\right)^{2 h-1} \label{sup-eq:current2} 
\\ &&\left| \Gamma 
  \left(  h +i \frac{\varepsilon}{2\pi T}\right)\right| ^2
  \left(P_0~e^{-\frac{\varepsilon}{2 T}}\right)  
  \left(\frac{\varepsilon}{2 T} \frac{\Delta T}{T} +\frac{eV}{2T}\right)\,,\nonumber 
\end{eqnarray} 
with $P_0$ the occupancy at equilibrium and here we set $T_R=T$.  This result coincides with the result of the main text (up to a proportionality factor). Hence, we show that our prediction of degeneracy induced asymmetry in the thermoelectric current is robust against presence of electron tunneling involving neutral charge non conserving processes.

\end{document}